\documentclass[prl,twocolumn,showpacs,floatfix,amsmath,amssymb,superscriptaddress]{revtex4-1}
\usepackage{amsfonts}
\usepackage{stmaryrd}
\usepackage{bbm}
\usepackage{mathrsfs}
\usepackage{tipa}
\usepackage{amssymb}
\usepackage{txfonts}
\usepackage{graphicx}
\usepackage{dcolumn}
\usepackage{epstopdf}
\usepackage[colorlinks,linkcolor=blue,urlcolor=blue,citecolor=blue]{hyperref}
\usepackage{multirow}
\usepackage{subfigure}
\usepackage{url}
\usepackage{upgreek}

\begin{document}
\newcommand*{\cm}{cm$^{-1}$\,}

\makeatletter
\newcommand{\Rmnum}[1]{\expandafter\@slowromancap\romannumeral #1@}
\makeatother

\title{Tunable near- to mid-infrared pump terahertz probe spectroscopy in reflection geometry}

\author{S. J. Zhang}
\affiliation{International Center for Quantum Materials, School of Physics, Peking University, Beijing 100871, China}

\author{Z. X. Wang}
\affiliation{International Center for Quantum Materials, School of Physics, Peking University, Beijing 100871, China}

\author{T. Dong}
\email{taodong@pku.edu.cn}
\affiliation{International Center for Quantum Materials, School of Physics, Peking University, Beijing 100871, China}

\author{N. L. Wang}
\email{nlwang@pku.edu.cn}
\affiliation{International Center for Quantum Materials, School of Physics, Peking University, Beijing 100871, China}
\affiliation{Collaborative Innovation Center of Quantum Matter, Beijing, China}

\begin{abstract}
Strong-field mid-infrared pump--terahertz (THz) probe spectroscopy has been proven as a powerful tool for light control of different orders in strongly correlated materials. We report the construction of an ultrafast broadband infrared pump--THz probe system in reflection geometry. A two-output optical parametric amplifier is used for generating mid-infrared pulses with GaSe as the nonlinear crystal. The setup is capable of pumping bulk materials at wavelengths ranging from 1.2 $\upmu$m to 15 $\upmu$m and beyond, and detecting the subtle, transient photoinduced changes in the reflected electric field of the THz probe at different temperatures. As a demonstration, we present 15 $\upmu$m pump--THz probe measurements of a bulk EuSbTe$_{3}$ single crystal. A $0.5\%$ transient change in the reflected THz electric field can be clearly resolved. The widely tuned pumping energy could be used in mode-selective excitation experiments and applied to many strongly correlated electron systems.

\end{abstract}


\maketitle
Over the past few decades, ultrafast spectroscopic techniques have provided new insights into exciting collective and single-particle modes of quantum materials and to track their subsequent decay pathways back to the equilibrium state. Most ultrafast experiments performed to date used near-infrared (NIR) and visible pulses, which are more accessible than mid-infrared (MIR) or far-infrared pulses, as most commercial ultrafast lasers are designed to generate narrowband pulses in the NIR range. In strongly correlated materials, in addition to the highly nonthermal electron distribution excited by high energy ($\sim$eV) photons \cite{PhysRevLett.65.2708,PhysRevLett.65.3445}, optical excitation can also result in a variety of novel dynamical phenomena, such as photoinduced insulator-to-metal transitions by ``photodoping'' Mott insulators \cite{PhysRevLett.91.057401} and optical melting of different orders \cite{PhysRevB.86.064425,Porer2014a}.

Low-energy excitations in strongly correlated materials, such as superconducting energy gap, Josephson plasma resonances, or specific lattice vibrations, always extend from gigahertz to MIR frequencies \cite{basov2011electrodynamics}. Thanks to recent developments in the generation of subpicosecond-duration laser pulses ranging from MIR to terahertz (THz) frequencies \cite{Hoffmann2011Intense}, low-energy excitations can be selectively controlled without delivering excess energy to other excitation pathways \cite{Fausti189,Dienst2013Optical,hu2014optically,Nicoletti2016}. Among such experiments, the most exciting one may be the observation of transient superconductivity in $YBa_2Cu_3O_{6.5}$ at room temperature \cite{Kaiser2014,hu2014optically}, in which the Josephson plasma edges formed by superfluid carriers in $YBa_2Cu_3O_{6.5}$ after excitation with 15 $\upmu$m pulses were detected in a $c$-axis time-domain THz measurement. Ultrafast broadband infrared pump--THz probe spectroscopy has been proven as a powerful tool for manipulating and detecting different orders in strongly correlated materials.

Here we report the construction of an ultrafast optical system capable of pumping a bulk sample from NIR to MIR frequencies (wavelengths of 1.2$\upmu$m--15$\upmu$m and even beyond) and probing at THz frequencies ($\sim$0.25--2.5 THz) in reflection geometry. We demonstrate the ability of this system to interrogate low-energy excitations in materials by photoexciting a bulk EuSbTe$_{3}$ crystal at room temperature using 15 $\upmu$m pulses and probing it with THz pulses, enabling us to time-resolve the photoinduced change in the THz electric field.

\begin{figure*}[htbp]
  \centering
  \includegraphics[width=15cm]{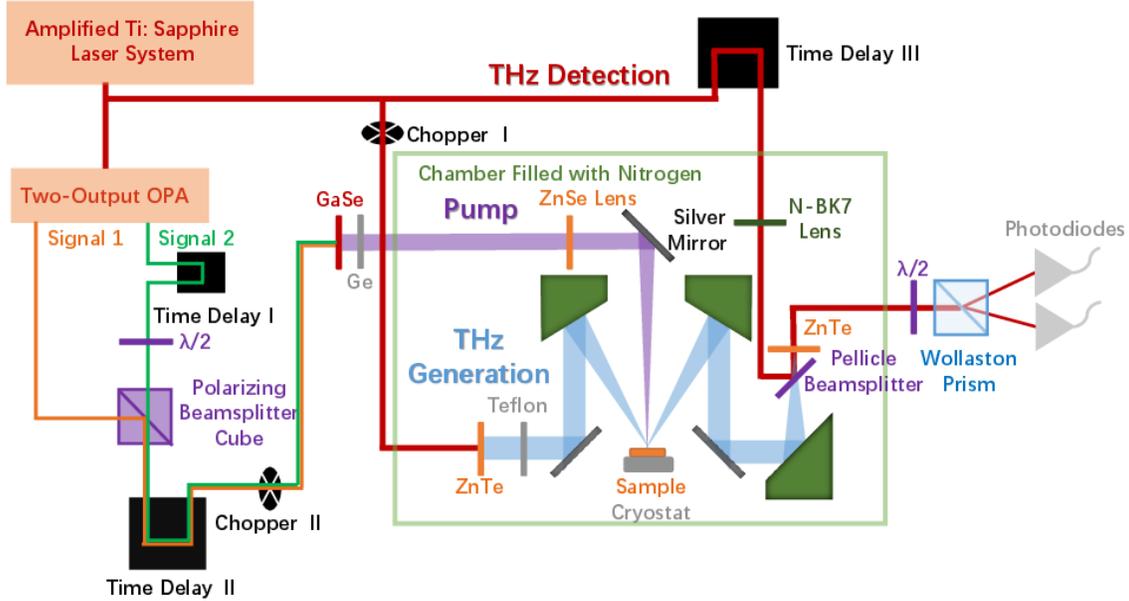}\\
  \caption{Experimental schematic for ultrafast broadband infrared pump--THz probe spectroscopy. The pump pulses enter the nitrogen chamber via a KBr window and are focused onto the sample at normal incidence. THz pulses are generated and detected in the chamber; 800 nm pulses, which enter the chamber via quartz windows, are used to generate THz pulses by optical rectification and detect the electric field of reflected THz by electro-optic sampling. THz pulses are focused onto the sample at 30${}^{\circ}$ incidence while the sample sits at the end of a cold finger in a cryostat.}\label{Fig:system}
\end{figure*}

Figure \ref{Fig:system} shows a diagram depicting the basic design of our optical system, which starts with an amplified Ti:sapphire laser system producing 800 nm, 35 fs pulses at a 1 kHz repetition rate from a regeneration amplifier. The system has three arms that are used for the pump, THz generation, and THz detection.

A two-output optical parametric amplifier (OPA) is used for pump pulse generation. The two outputs both contain signal and idler beams ranging from 1.2 to 1.6 $\upmu$m and 1.6 to 2.6 $\upmu$m, respectively. Both the signal and idler beams can be used as the NIR pump in spectroscopy without focalization because they have sufficiently high fluences. To obtain the MIR pump, two signal beams that have been tuned to perpendicular polarization are used for difference frequency generation (DFG) collinearly on a 1-mm-thick z-cut GaSe crystal \cite{Vodopyanov2002}. By tuning the frequency of the two signal beams and the orientation of the GaSe crystal to meet the type \Rmnum{1} or type \Rmnum{2} phase-matching condition, MIR pulses with tunable polarization ranging from 3 to 15 $\upmu$m can be generated \cite{huber2000generation,reimann2003direct}. A 5-mm-thick germanium plate used for blocking the two signal pulses is placed after the GaSe crystal. Pump pulses of 15 $\upmu$m, where the available pump power is minimum in the current measurement, carry up to 2.5 $\upmu$J per pulse. A ZnSe convex lens is used to focus the pump pulses generated by DFG to obtain sufficient pump fluence.
\begin{figure}[htbp]
  \centering
  \includegraphics[width=8.5cm]{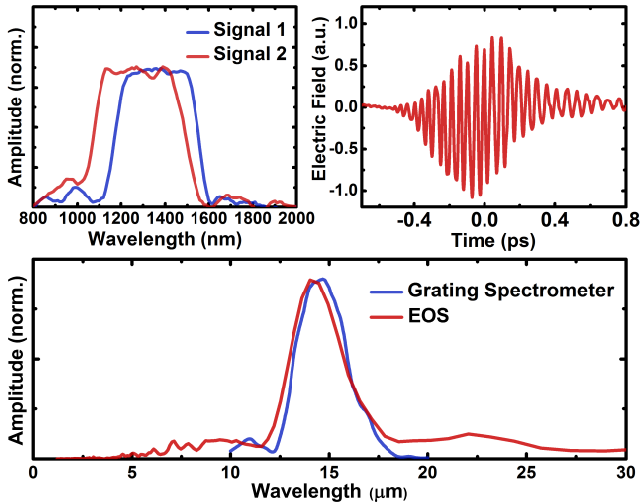}\\
  \caption{Top left: Frequency spectra of two signal pulses obtained by a grating spectrometer. Top right: Time domain electric field of MIR pulses generated via DFG and measured via EOS in air. Bottom: Frequency spectra of MIR pulses obtained by EOS and a grating spectrometer. }\label{Fig:mir}
\end{figure}

The two signal beams used for DFG are seeded by the same white light continuum. So the generated MIR pulses are intrinsically carrier envelope phase stable, and their electric fields can be measured via electro-optic sampling (EOS) \cite{Huber2008} with 35 fs gate pulses focused on a 100-$\upmu$m-thick z-cut GaSe crystal \cite{liu2004gase,kubler2004ultrabroadband}. To obtain MIR pulses centered at 15 $\upmu$m, the two signals of the OPA should be set at the frequencies shown in the top left panel of Fig.\ref{Fig:mir}, which are measured with a grating spectrometer. The time-domain electric field of the 15 $\upmu$m pulses generated via DFG are measured via EOS in air, as shown in the top right panel of Fig.\ref{Fig:mir}. The frequency-domain spectrum of the difference frequency pulses is acquired by Fourier transformation of the EOS signal, which is shown in the bottom panel of Fig.\ref{Fig:mir}. We also measure the spectrum of the difference frequency pulse directly with a grating spectrometer equipped with a liquid-nitrogen-cooled mercury cadmium telluride detector. The result is also presented in the bottom panel of Fig.\ref{Fig:mir} for comparison. The two different measurements show very good agreement. The subtle differences between the two frequency spectra may result from the large susceptibilities to the orientation of the GaSe crystal and the non-negligible bandwidth of the two signal pulses.

\begin{figure}[htbp]
  \centering
  \includegraphics[width=8.5cm]{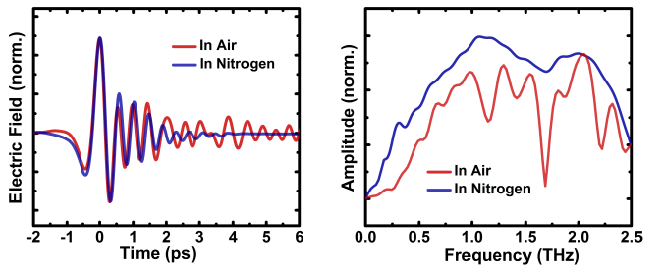}\\
  \caption{Time-dependent electric field of THz probe pulse (left) and corresponding Fourier amplitude as a function of frequency (right) measured in air and nitrogen. }\label{Fig:static}
\end{figure}

The THz probe pulses are generated from pulses of 800 nm light whose polarization can be tuned by a $\lambda/2$-plate using a 1-mm-thick (110) ZnTe crystal \cite{Rice1994Terahertz} without focalization. By using a 30${}^{\circ}$ off-axis parabolic mirror whose focal length is 54.45 mm, we obtain a 0.63 mm THz spot size, as measured with a knife-edge scan. This rather small THz spot allows us to focus the pump beam into a spot less than 0.7 mm in size, resulting in generation of a higher MIR pump fluence. The THz profile was detected via EOS using 1-mm-thick ZnTe as the nonlinear detection crystal \cite{Wu1997Free}. The THz pulses and 800 nm sampling pulses are focused onto the ZnTe crystal at normal incidence using a 90${}^{\circ}$ off-axis parabolic mirror and an N-BK7 convex lens, respectively. Figure \ref{Fig:static} shows the time-dependent electric field of the THz pulse measured in air and nitrogen and the corresponding spectra as a function of frequency, obtained via Fourier transformation, of a test sample of EuSbTe$_{3}$ \cite{Niu2015b} at room temperature. The dips in the Fourier amplitude spectra can be attributed to water absorption in air, and their positions in the frequency domain agree well with previously reported results \cite{Fattinger1989Terahertz,Yang:11}. The water absorption could be eliminated by either evacuating the chamber in which the THz beam is generated or filling the chamber with dry nitrogen gas. Because some mirrors in the chamber could undergo subtle changes in direction/position during evacuation, the current measurement was performed by filling the chamber with nitrogen gas. Samples sit at the end of a cold finger in a helium continuous-flow cryostat, which is capable of reaching a base pressure as low as 2$\times10^{-5}$ Pa and a temperature of 4 K.

There are three time delays in this system, as shown in Fig.\ref{Fig:system}, which are set along the signal 2 path (time delay \Rmnum{1}), pump arm (time delay \Rmnum{2}), and THz detection arm (time delay \Rmnum{3}). Time delay \Rmnum{1} is used for precise compensation of the optical path differences between signal 1 and signal 2, which is a prerequisite for DFG. By moving time delay \Rmnum{2} while keeping time delay \Rmnum{3} at the peak position of the THz electric field, the pump--probe time delay, denoted as $\tau$, can be acquired. $\tau$ refers to the delay after excitation of the sample and reveals the changes in the peak of the THz electric field induced by a pump pulse. The transient changes in the time-dependent THz electric field at $\tau$ can be captured by moving both time delay \Rmnum{2} and time delay \Rmnum{3} in the same step to set the relative delay between the detection pulses and THz pulses while keeping the pump and detection pulses relatively still. The time delay between the sampling pulse and the THz pulse, $t$, maps out the THz spectrum at each $\tau$. The time resolution of this system is not determined by the envelope of the THz pulses, which is generally over 1 ps long, but by the Fourier limit of the pulse, \emph{i.e.}, the pulse width of 35fs.

There are two choppers in this system, as shown in Fig.\ref{Fig:system}, which are set along the THz detection arm (chopper \Rmnum{1}) and the pump arm (chopper \Rmnum{2}). There are two methods of acquiring the transient changes in the THz electric field $\Delta E(t)$. In one case, the equilibrium reflected electric field $E_0(t)$ and the after-pumping field $E'(t)$ are measured by modulation using chopper \Rmnum{1}, so $\Delta E(t)$ can be calculated as $\Delta E(t) = E'(t) - E_0(t)$. In the other case, $\Delta E(t)$ is acquired directly by modulation of the pump pulse with chopper \Rmnum{2} by filtering the EOS signal with a lock-in amplifier. Typically, the results of these two methods should be identical, which was confirmed with our spectroscopy by the measurement of EuSbTe$_{3}$ as shown below.

To evaluate the performance of our system, we performed 15 $\upmu$m pump and 1.2 $\upmu$m pump--THz probe measurements of a bulk EuSbTe$_{3}$ crystal at room temperature.

\begin{figure}[htbp]
  \centering
  \includegraphics[width=8.5cm]{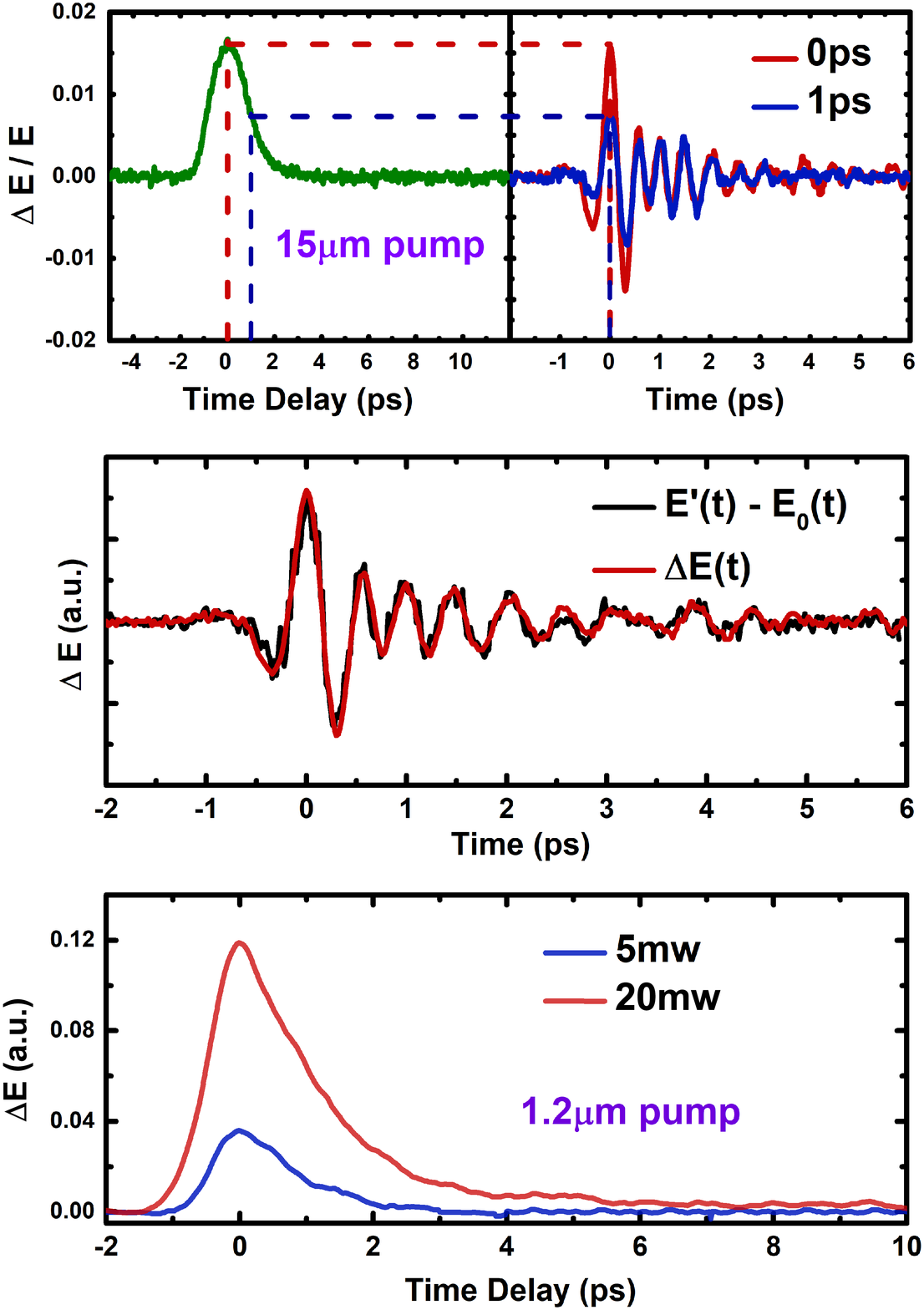}\\
  \caption{15 $\upmu$m and 1.2 $\upmu$m pump--THz probe measurements of EuSbTe$_{3}$ at room temperature. Top left: Relaxation of the sample after 15 $\upmu$m excitation. Top right: Pump-induced changes in the reflected THz electric field ($\Delta E/E$) at each $\tau$. Center: Difference between the equilibrium reflected electric field $E_0(t)$ and the after-pumping field $E'(t)$ at $\tau = 0$ ps (red), and $\Delta E(t)$ measured directly by modulation of the pump (blue). Bottom: Relaxation of the sample after 1.2 $\upmu$m excitation.}\label{Fig:Eu}
\end{figure}

The top left panel in Fig.\ref{Fig:Eu} depicts the relaxation of the sample after excitation by 15 $\upmu$m pump pulses. Here pump scans are taken by fixing THz time delay III at the peak position of the THz electric field. The pump-induced changes in the reflected THz electric field ($\Delta E/E$) acquired directly by modulation of the pump pulse with chopper \Rmnum{2} at each $\tau$ are shown in the top right panel in Fig.\ref{Fig:Eu}, in which the changes in the electric fields match well with those observed in the pump--probe measurements (dashed lines are guides for the eyes.). A $0.5\%$ transient change in the reflected THz electric field can be unambiguously measured in our experiment.

For EuSbTe$_{3}$, the equilibrium reflected electric field $E_0(t)$ and the after-pumping field $E'(t)$ at $\tau = 0$ ps were also measured by modulation with chopper \Rmnum{1}, and $\Delta E(t)$ was calculated using the function $\Delta E(t) = E'(t) - E_0(t)$. These two different methods are compared in the center panel of Fig.\ref{Fig:Eu}. It can be concluded that the obtained transient changes in the THz reflected electric field ($\Delta E/E$) are identical regardless of which pump pulses or THz pulses are modulated. However, a much higher signal-to-noise ratio can be achieved by modulation of the pump pulse because the pump-induced changes in the field $\Delta E(t)$ are very small compared to $E_0(t)$. 
The 15 $\upmu$m pump--THz probe experimental measurement of bulk EuSbTe$_{3}$ demonstrates that the constructed ultrafast broadband infrared pump--THz probe spectroscopy is credible and sensitive enough to observe subtle changes in the reflectivity configuration. It is expected to reveal novel phenomena in time-resolved THz spectroscopy experiments, in particular in mode-selective control measurement of strongly correlated electron systems.

NIR pump--THz probe spectroscopy can be enabled by simply blocking one of the signal beams from the two-output OPA and removing the germanium plate in the pump arm, as a slight displacement of the temporal and spatial overlaps of the pump and probe pulses can be easily corrected by moving time delay \Rmnum{2} and tuning the adjuster screws of the silver mirror in Fig.\ref{Fig:system}, respectively. In fact, NIR pump--THz probe spectroscopy is much easier to manipulate than MIR pump spectroscopy, as there is always a small amount of visible light in the NIR pulses that can be used for spatially overlapping the pump and THz-generating 800 nm pulses, \emph{i.e.}, overlapping the pump and THz probe pulses by direct visual observation (the subtle displacement between the THz pulses and the 800 nm pulses is negligible here). The bottom panel of Fig.\ref{Fig:Eu} shows the relaxation of the sample after excitation by 1.2 $\upmu$m pump pulses with two different fluences. The pump-induced change in the THz electric field at each $\tau$ could be obtained as described for the 15 $\upmu$m pump case. This experimental apparatus can also be used for a 800 nm pump--THz probe experiment by compensating for the optical path differences. A 400 nm pump--THz probe experiment is feasible using a barium borate crystal.


To summarize, we constructed an ultrafast broadband infrared pump--THz probe spectroscopy system, which is capable of detecting subtle changes in the reflected electric field of the THz probe. A two-output OPA is used for generating MIR pulses with wavelengths of 15 $\upmu$m or beyond, where GaSe is used as the nonlinear crystal. This type of ultrafast broadband spectroscopy has been proven as a powerful tool for light control of different orders in strongly correlated materials, which is able to pumping low energy excitations in materials and probing the transient photoinduced changes with terahertz pulses. Room-temperature measurements of bulk EuSbTe$_{3}$ using a 15 $\upmu$m pump and 0.2--2.5 THz probe were performed, and a $0.5\%$ transient change in the reflected THz electric field was clearly detected with sufficient signal-to-noise ratio.

\begin{center}
\small{\textbf{ACKNOWLEDGMENTS}}
\end{center}
We thank Dr. Dong Wu for supplying the EuSbTe$_{3}$ single crystals that were used for the apparatus testing measurements.
This work was supported by the National Science Foundation of China (No. 11327806, GZ1123) and the National Key Research and Development Program of China (No. 2016YFA0300902).

\bibliographystyle{apsrev4-1}
  \bibliography{Terahertz}

\end{document}